# TWO ANSWERS TO A COMMON QUESTION ON DIAGONALIZATION

Samuel C. Hsieh
Computer Science Department, Ball State University, Muncie, Indiana, U.S.A.

**ABSTRACT**
A common question from students on the usual diagonalization proof for the uncountability of the set of real numbers is: when a representation of real numbers, such as the decimal expansions of real numbers, allows us to use the diagonalization argument to prove that the set of real numbers is uncountable, why can't we similarly apply the diagonalization argument to rational numbers in the same representation? why doesn't the argument similarly prove that the set of rational numbers is uncountable too? We consider two answers to this question. We first discuss an answer that is based on the familiar decimal expansions. We then present an unconventional answer that is based on continued fractions.

## 1. INTRODUCTION

In discrete mathematics and theory of computation classes, the set of real numbers, henceforth denoted by **R**, is commonly shown to be uncountable by a contradiction proof using the diagonalization argument, e.g., in [1, 2]. For the purpose of contradiction, such a proof assumes that **R** is countable and thus there is a one-to-one correspondence between the set of natural numbers and **R**. The assumption is then contradicted by showing the existence of a real number that is not included in the one-to-one correspondence. More specifically, by the assumption that **R** is countable, **all** of the real numbers in **R** can be enumerated in a countably infinite sequence

$$L = r_1, r_2, r_3, \ldots,$$

where each $r_i$ is a real number in **R**. Let each real number $r_i$ be represented by its decimal expansion and let $d_{ij}$ denote the digit in the $j^{th}$ decimal place of the decimal expansion of $r_i$. That is, each real number $r_i$ in the sequence L is represented by the following expression:

$$w_i . d_{i1}\ d_{i2}\ d_{i3}\ \ldots$$

where $w_i$ is the integer portion of $r_i$, and $d_{i1}\ d_{i2}\ d_{i3}\ldots$ is an infinite sequence of digits that represents the fractional part of $r_i$ and possibly includes trailing 0's. The fractional parts of all of the real numbers $r_1, r_2, r_3, \ldots$ in the sequence L, that is, the digits $d_{ij}$ for all $i \geq 1$ and for all $j \geq 1$, form a matrix as shown below in Fig. 1.

$$r_1 = w_1.\mathbf{d_{11}}\ d_{12}\ d_{13}\ d_{14}\ d_{15}\ d_{16}\ldots$$
$$r_2 = w_2.d_{21}\ \mathbf{d_{22}}\ d_{23}\ d_{24}\ d_{25}\ d_{26}\ldots$$
$$r_3 = w_3.d_{31}\ d_{32}\ \mathbf{d_{33}}\ d_{34}\ d_{35}\ d_{36}\ldots$$
$$r_4 = w_4.d_{41}\ d_{42}\ d_{43}\ \mathbf{d_{44}}\ d_{45}\ d_{46}\ldots$$
$$r_5 = w_5.d_{51}\ d_{52}\ d_{53}\ d_{54}\ \mathbf{d_{55}}\ d_{56}\ldots$$
$$r_6 = w_6.d_{61}\ d_{62}\ d_{63}\ d_{64}\ d_{65}\ \mathbf{d_{66}}\ldots$$
$$\ldots$$
$$\ldots$$
$$\ldots$$

Fig. 1 Matrix and Diagonal

The diagonal of the matrix is the sequence of digits $d_{kk}$ for all k=1,2,3,…, as shown in bold face in Fig. 1. From the diagonal, a *new* number

$$r_0 = w_0.d_{01}\ d_{02}\ d_{03}\ \ldots$$

can be derived such that for all k=1, 2, 3 …, $d_{0k}$ differs from $d_{kk}$ and is neither 0 nor 9. In other words, the new number $r_0$ differs from the real number $r_1$ in the 1$^{st}$ decimal place, from $r_2$ in the 2$^{nd}$ decimal place and, in general, from each real number $r_k$ in the k$^{th}$ decimal place. Although a real number can have an alternative representation in decimal form, for example 5.26999999… is the alternative decimal representation of 5.27000000… and vice versa, $r_0$ cannot be the alternative representation of any real number since the digits $d_{01}\ d_{02}\ d_{03}\ \ldots$ can neither be 0 nor 9. Therefore, the new number $r_0$ cannot be equal to any real number in the sequence L. This contradicts the assumption that the sequence L includes all real numbers in **R**. Hence, **R** is not countable. This argument is based on the existence of the new number $r_0$, which is derived from the diagonal of the matrix in Fig. 1, hence the name *diagonalization argument*.

When such a proof is discussed in discrete mathematics or theory of computation classes, a question that often arises is as follows. When a representation of real numbers allows us to use the diagonalization argument to prove that the set of real numbers is uncountable, can we not apply the argument similarly to rational numbers in the same representation? Why doesn't the diagonalization argument similarly prove that the set of rational numbers is uncountable then? For example, when rational numbers are represented by their decimal expansions, why can't we similarly apply the diagonalization argument to rational numbers and prove that the set of rational numbers is uncountable, as we do in the case of real numbers?

The next section discusses an answer that is based on the familiar decimal expansions, and section 3 provides an unconventional answer based on continued fractions. Section 3 also includes a brief introduction to continued fractions. For the sake of presentation, in this paper when we refer to **positive** real, rational, or irrational numbers, we often omit the word positive.

## 2. AN ANSWER BASED ON DECIMAL EXPANSIONS

Suppose that we wish to apply the diagonalization argument to rational numbers. The set of rational numbers can of course be enumerated in some sequence $f_1, f_2, f_3\ldots$ with each rational number $f_i$ represented by its decimal expansion $w_i.d_{i1}\ d_{i2}\ d_{i3}\ \ldots$, possibly with trailing 0's. The digits $d_{ij}$ for all i ≥ 1 and for all j ≥ 1 would then form a matrix similar to that shown in Fig. 1. For the purpose of applying the diagonalization argument, a *new* number $f_0$ can then be derived from the diagonal of this matrix, such that $f_0$ differs from every rational number in the enumeration $f_1, f_2, f_3\ldots$. This leads to an answer to the question raised in the previous section: this new number $f_0$ is **not** rational and, hence, the diagonalization argument does not prove that the set of rational numbers is uncountable. This answer, though concise, does require elaboration as to why the new number $f_0$ is not rational.

It is well known that the decimal expansion of any rational number, after a number of decimal places, infinitely repeats some finite sequence of digits. For example, 1/6 is 0.1 6 6 6… and so on, and 169/550 is 0.23 45 45 45… and so on. Such a decimal expansion is said to be *periodic*. The repeated sequence in a periodic decimal expansion is known as its *period*, such as the sequence 6 in 0.1 6 6 6… and the sequence 45 in 0.23 45 45 45… . The number of digits in a period is its *period length*. For example, the period length of 0.1 6 6 6… is 1 and that of 0.23 45 45 45… is 2. The new number $f_0$, which is derived from the diagonal of a matrix formed by the digits of the fractional parts of all rational numbers in the enumeration $f_1$, $f_2$, $f_3$…, is not rational because it does not have a period length.

Naturally, the question as to why the new number $f_0$ does not have a period length will need to be addressed. Since every rational number has a finite period, it is natural to question why the new number $f_0$ can not have, as its period length, the least common multiple of the period lengths of all rational numbers. An answer is that although every rational number has a finite period length, there is not an upper bound on the period lengths of rational numbers. Hence, there is not an upper bound on the least common multiple of the period lengths of all rational numbers.

### 3. AN UNCONVENTIONAL ANSWER

This section presents another answer to the question raised in section 1. Since this answer uses continued fractions, a brief introduction to continued fractions is given in section 3.1, and the answer itself is provided in section 3.2.

### 3.1 Continued Fractions

Any positive real number, rational or irrational, can be represented in the following *staircase notation* (Fig. 2), where $a_0$ is a nonnegative integer and $a_1$, $a_2$, $a_3$, $a_4$ … are positive integers. This representation of a real number is called a *continued fraction*.

$$a_0 + \cfrac{1}{a_1 + \cfrac{1}{a_2 + \cfrac{1}{a_3 + \cfrac{1}{a_4 \cdots}}}}$$

Fig. 2 Staircase Notation

For ease of presentation, henceforth a list representation will be used instead of the staircase notation. In the list representation, the above continued fraction is written as [$a_0$; $a_1$, $a_2$, $a_3$, $a_4$ …]. Hence, as a continued fraction, any positive real number can be represented by a sequence of nonnegative integers. For example, 6/7 is [0; 1, 6] as a continued fraction and 0.857142 857142… in decimal form; and the square root of 2 (an irrational number) is [1;2,2,2,2,2…] as a continued fraction and 1.414213562373095… in decimal form. It is interesting to note the regularity that is present in the continued fraction [1;2,2,2,2,2…] and the lack of regularity in the decimal expansion 1.414213562373095…. of the same number.

Algorithms for converting a rational number or an irrational number to an equivalent continued fraction are straightforward and are presented below as static methods in Java.

```java
public static String cf(int num, int den)
  { String result="";
    int temp;

    do {
        result=result+num / den + " ";
        temp=num % den;
        num=den;
        den=temp;
      } while(den != 0);
    return result;
  }  //end cf
```

The method `cf` in the above Java code implements an algorithm to convert a positive rational number to an equivalent continued fraction. Given two positive integers as the numerator (`num`) and the denominator (`den`) of a fraction representing a rational number, the method returns the equivalent continued fraction in the list representation - as a string consisting of a list of space-separated integers. For example, `cf(6,7)` returns the string "0 1 6". The algorithm is similar to the Euclidean algorithm for finding the greatest common divisor and terminates similarly. In other words, every rational number can be represented by a finite list of integers as a continued fraction.

```java
public static String cfIr(double x)
  { String result="";
    double temp=x;

    do {
        result=result + (int) temp+" ";
        temp=1.0/(temp - (int) temp);
     } while (Math.abs(x-eval(new Scanner(result)))>EPS);

     return result;
   }
```

The method `cfIr` in the above Java code implements a similar algorithm to convert a positive irrational number (approximated as a `double` value in the above Java code) to its equivalent continued fraction. Given an irrational number, the method returns an (approximate) continued fraction as a list. For example, the method call `cfIR(MATH.sqrt(2))` (to find the continued fraction of the square root of 2) returns the list "1 2 2 2 2...". In the above code, the method call `eval(new Scanner(result))` returns the value of the continued fraction represented by the string parameter `result`, and `EPS` is a constant, usually a very small value, that specifies the accuracy of the resultant continued fraction, as described below. Every irrational number, when represented as a continued fraction, is an infinite sequence of integers. Of course, the method `cfIr` can only return a finite sequence that approximates a given irrational number (which is also approximated as a `double` in the above Java code). The constant `EPS` specifies how closely the returned continued fraction should approximate the given irrational number – a small `EPS` leads to a close approximation and a long list representing the continued fraction. The method `eval` implements an algorithm that is straightforward, as is the Java code for `eval`, which is not included here as the code for the method `eval` is not essential to an understanding of the algorithms presented above.

**3.2 An Answer Based on Continued Fractions**

We now provide a continued fraction-based answer to the question raised in section 1: when a representation of real numbers allows us to use the diagonalization argument to prove that the set of real numbers is uncountable, can we not apply the argument similarly to rational numbers in the same representation? Why doesn't the diagonalization argument similarly prove that the set of rational numbers is uncountable then? Indeed, the continued fraction representation allows us to use the diagonalization argument to prove that the set of irrational numbers, and hence the set of real numbers, is uncountable, as outlined below. For the purpose of contradiction, let us assume that the set of irrational numbers can be enumerated in some sequence $r_1$, $r_2$, $r_3$… with each irrational number $r_i$ represented as a continued fraction by an infinite list [$a_{i0}$; $a_{i1}$, $a_{i2}$, $a_{i3}$, …], where $a_{ik}$ for each $k \geq 0$ is the $k^{th}$ integer in the list representation of the irrational number $r_i$. The integers $a_{ik}$ for all $i \geq 1$ and for all $k \geq 1$ can then form a matrix of integers, similar to that shown in Fig. 1, with the sequence $a_{11}$, $a_{22}$, $a_{33}$…, that is, the sequence $a_{kk}$ for all $k \geq 1$, as the diagonal of this matrix. A *new* irrational number $r_0 = $ [$a_{00}$; $a_{01}$, $a_{02}$, $a_{03}$…], where $a_{00}$ is any non-negative integer, can be derived from the diagonal of this matrix such that for all $k \geq 1$, $a_{0k}$ differs from $a_{kk}$ and is not 0. That is, $r_0$ differs from every irrational number in the enumeration $r_1$, $r_2$, $r_3$… . This contradicts the assumption that the set of irrational numbers can be enumerated in a sequence. Therefore, the set of irrational numbers, and hence the set of real numbers, is uncountable.

Similar to the proof in section 1, this diagonalization proof depends on the existence of a new number $r_0 = $ [$a_{i0}$; $a_{01}$, $a_{02}$, $a_{03}$, …] that is derived from the countably **infinite** diagonal $a_{11}$, $a_{22}$, $a_{33}$ … such that $a_{01}$ differs from $a_{11}$, $a_{02}$ from $a_{22}$, $a_{03}$ from $a_{33}$, and so on. While such a diagonal exists in any countably infinite sequence of irrational numbers represented as continued fractions, such a countably **infinite** diagonal **does not** exist in any countably infinite sequence of rational numbers represented in the form of continued fractions, because, as a continued fraction, every rational number is represented by a **finite** list of integers. Hence, the diagonalization argument cannot be applied to the set of rational numbers in the continued fraction representation,

Besides this salient feature that as a continued fraction, every rational number is represented by a finite list and every irrational number is represented by an infinite list, continued fractions are interesting for other reasons, such as the two given below, and should be an interesting topic to introduce briefly in a discrete mathematics class or a related class.

- Continued fractions often clearly reveal the structures of irrational numbers: for example, the square root of 2 is [1;2,2,2,…] as a continued fraction but 1.414213562373095… in decimal form, and the number *e* is [2;1,2,1,1,4,1,1,6,1,1,8…] as a continued fraction but 2.718281828459… in decimal form.
- Continued fractions provide a good approximation of irrational numbers, for example, the continued fraction [3; 7, 15, 1], which uses 4 digits to represent the fractional part, is 3.141592…, which is a much closer approximation of PI than 3.1416, which also uses 4 digits to represent the fractional part.